\documentclass[preprint2]{aastex}  
\usepackage{rotating}
\usepackage{lineno}
\pdfoutput=1

\shorttitle{SUPERNOVA SEARCH AT LVD}
\shortauthors{AGAFANOVA ET AL.}

\begin{document}

\title{Implication for the core collapse supernova rate from 21 years of data of the
Large Volume Detector}


\author{
N.Y.Agafonova\altaffilmark{1}, 
M.Aglietta\altaffilmark{2}, 
P.Antonioli\altaffilmark{3}, 
V.V.Ashikhmin\altaffilmark{1},
G.Badino\altaffilmark{2,7},
G.Bari\altaffilmark{3}, 
R.Bertoni\altaffilmark{2}, 
E.Bressan\altaffilmark{4,5}, 
G.Bruno\altaffilmark{6}, 
V.L.Dadykin\altaffilmark{1}, 
E.A. Dobrynina\altaffilmark{1}, 
R.I.Enikeev\altaffilmark{1}, 
W.Fulgione\altaffilmark{2,6,*}, 
P.Galeotti\altaffilmark{2,7}, 
M.Garbini\altaffilmark{3}, 
P.L.Ghia\altaffilmark{8}, 
P.Giusti\altaffilmark{3}, 
F.Gomez\altaffilmark{2},
E.Kemp\altaffilmark{9},  
A.S.Malgin\altaffilmark{1}, 
A.Molinario\altaffilmark{2,6}, 
R.Persiani\altaffilmark{3}, 
I.A.Pless\altaffilmark{10}, 
A.Porta\altaffilmark{2}, 
V.G.Ryasny\altaffilmark{1}, 
O.G.Ryazhskaya\altaffilmark{1}, 
O.Saavedra\altaffilmark{2,7}, 
G.Sartorelli\altaffilmark{3,4}, 
I.R.Shakiryanova\altaffilmark{1},
M.Selvi\altaffilmark{3}, 
G.C.Trinchero\altaffilmark{2},
C.Vigorito\altaffilmark{2,7,*}, 
V.F.Yakushev\altaffilmark{1}, 
A.Zichichi\altaffilmark{3,4,5,11} \\
(The LVD Collaboration)}

\altaffiltext{*} {Corresponding authors: \\fulgione@to.infn.it and vigorito@to.infn.it}
\altaffiltext{1} {Institute for Nuclear Research, Russian Academy of Sciences, Moscow, Russia}
\altaffiltext{2} {INFN-Torino, OATo-Torino, 10100 Torino, Italy}
\altaffiltext{3} {INFN-Bologna, 40126 Bologna, Italy} 
\altaffiltext{4} {University of Bologna, 40126 Bologna, Italy}
\altaffiltext{5} {Centro Enrico Fermi, 00184 Roma, Italy}
\altaffiltext{6} {INFN-Laboratori Nazionali del Gran Sasso and Gran Sasso Science Institute, 67100 LÕAquila, Italy}
\altaffiltext{7} {Dep.of Physics, University of Torino, 10125 Torino, Italy}
\altaffiltext{8} {CNRS-IN2P3, Paris, Laboratoire de Physique Nucl\'{e}aire et de Hautes Energies (LPNHE), Universit\'{e}s Paris 6 et Paris 7, France}
\altaffiltext{9} {University of Campinas, 13083-859 Campinas, SP, Brazil}
\altaffiltext{10} {Massachusetts Institute of Technology, Cambridge, MA 02139-4307, USA}
\altaffiltext{11} {CERN, Geneva, Switzerland}

\begin{abstract}
The Large Volume Detector (LVD) has been continuously taking data since 1992 at the INFN Gran Sasso National Laboratory. LVD is sensitive to neutrino bursts from gravitational stellar collapses with full detection probability over the Galaxy. We have searched for neutrino bursts in LVD data taken in 7335 days of operation. No evidence of neutrino signals has been found between June 1992 and December 2013. The 90\% C.L. upper limit on the rate of core-collapse and failed supernova explosions out to distances of 25 kpc is found to be 0.114 y$^{-1}$ .
\end{abstract}




\section{Introduction}
\label{intro}

The detection of neutrinos from the optically bright supernova in the Large Magellanic Cloud, SN1987A \citep{1987Hirata, 1987Bionta, 1987Alekseev} and \citep{1987Aglietta}\footnote{The explanation of five signals recorded by the LSD detector about 5 hours earlier with respect to the other three experiment still remains controversial.}, led to important inferences on the physics of core collapse supernovae. It experimentally proved the critical role of neutrinos in the explosion of massive stars, as suggested more than 50 years ago \citep{1940Gamow},\citep{1965Zel'dovich},\citep{1966Colgate},\citep{1977Nadyozhin}. While a complete understanding of the physics involved is still lacking (see e.g. \citep{Woosley 2005}) the SN1987A event helped to establish some aspects of the theory, namely the total energy radiated, the neutrinos temperatures and the duration of the radiation pulse (see e.g. \citep{2002Loredo} \citep{2009Pagliaroli_a}).

However, only a small number of neutrinos could be detected in that occasion, $\approx 20$. 
Thus, it was not possible to study the detailed features of the neutrino emission, which is expected to carry important information on the dynamics of the explosion.
Such a small number was due not only to the source distance (about 50 kpc from the Earth) but also to the relatively small dimensions of the detectors existing at that time. In fact, the need for larger and more sensitive neutrino detectors to study one of the most powerful and rare events occurring in the Galaxy had already become evident in the scientific community even before SN 1987A. The extremely low frequency (present estimates give a rate between one every 10 yr and one every 100 yr) implies that long-term observations using powerful neutrino detectors are of essence to detect explosions of massive stars.
Also, the observation of neutrinos from SN 1987A was guided by the optical observation. However, the core-collapse rate in the Galaxy exceeds that of observable optical supernovae because light can be partially or totally absorbed by dust in the Galactic plane. In recent times this point has been discussed by \citep{2013Adams} with the conclusion that large long-term neutrino detectors are the most suited to observing the Galaxy searching for core- collapse supernovae explosions. Neutrino detectors are also sensitive to collapsing objects that fail to explode, becoming black holes (so-called failed supernovae), because those are expected to emit a neutrino signal even stronger, although shorter in time, than from core- collapse supernovae \citep{2008Nakazato}.

In addition, the prompt identification of a neutrino signal could provide astronomers with an early alert of a supernova occurrence (SuperNova Early Warning System, SNEWS, \citep{2004Antonioli} of which LVD is a founding member) allowing one to study phenomenons like the shock break out, a flash of radiation as the shock wave breaks out from the surface of the star \citep{1978Klein} \citep{1978Falk}, and to detect, for the first time directly, the signal due to the emission of gravitational waves \citep{2009Pagliaroli_b}.

Based on the pioner idea by \citep{1965Zaz}, several neutrino detectors have been observing the Galaxy in the last decades to search for stellar collapses, namely Super-Kamiokande \citep{2007SuperK}, Baksan \citep{2011Novoseltseva}, MACRO \citep{2004Ambrosio}, AMANDA \citep{2002Ahrens}, SNO \citep{2011Aharmim}. None of them has found evidence of supernovae explosions, thus setting limits to the rate of collapses. The longest duration experiment is Baksan: it has provided the most stringent limit in terms of rate (0.09 per year at 90\% C.L. based on 26 years of operation) but given the limited size its sensitivity to the whole Galaxy is controversial. In turn, the most sensible detector, Super-Kamiokande (fully efficient up to 100 kpc), sets a limit to the rate of 0.32 per year at 90\% C.L.

In this paper we present the results of the search for supernova neutrino bursts based on the data taken by the Large Volume Detector (LVD) \citep{1992Aglietta} in more than 20 yr of operation in the INFN Gran Sasso National Laboratory (LNGS) in Italy. The concept of a powerful neutrino detector was actually a basic motivation of the LNGS project itself started at the end of the 70s (see \citep{1999Bettini}; \citep{2000Zichichi} for a historical review). The LVD is a large-mass (1,000 t), long-term (operating since 1992) neutrino experiment located at the depth of 3,600 m w.e..

The detector main characteristics are described in Section \ref{Detector}. The data set used in this work extends from June 1992 to December 2013. In this period, LVD has recorded more than 5 billions of triggers, mostly due to radioactive background and atmospheric muons. In Section \ref{sec:dataanalysis} we explain the criteria for reducing such backgrounds and selecting events potentially due to neutrinos. To search for supernova neutrino bursts, we analyze the time series of those events and search for clusters. While to provide the SNEWS with a prompt alert we use in the burst-search algorithm a fixed-time window (20 s)\citep{2012Agafonova}, in this work we consider different burst durations up to 100 s. The analysis is detailed in Section \ref{search}. In the same section we also discuss the sensitivity of the analysis to the recognition of a supernova event by using a conservative model based on the observations of neutrinos from the SN1987A\citep{2009Pagliaroli_a}. 
Finally, in Section \ref{results}, we present the results of the search for neutrinos from gravitational stellar collapses happening in the whole Galaxy. Our conclusions are given in  \ref{conclusions}.
Finally, we present the results of the search for neutrinos from gravitational stellar collapses happening in the whole Galaxy in Section \ref{results} and conclusions are given in Section \ref{conclusions}.

\section{The Large Volume Detector}
\label{Detector}

\begin{deluxetable}{cccc}
\tabletypesize{\scriptsize}
\tablecaption{$\nu$ interaction channels in LVD. Cross sections of different interactions are obtained referring to \citep{updStrumia} for interaction 1, \citep{1988Fukugita} for interactions 2-4, \citep{1995Bahcall} for interaction 5 and \citep{2001Kolbe} and \citep{2001Toivanen} for interactions 6-8.\label{tab:interactions}}
\tablewidth{0pt}
\tablehead{
\colhead{} & \colhead{$\nu$ interaction channel} & \colhead{$\mathrm{E_{\nu}}$ threshold} & \colhead{$\mathrm{\%}$}
}
\startdata
1 & $\mathrm{\bar \nu_\mathrm{e} + \mathrm{p \rightarrow e^{+} + n}}$ & (1.8 MeV) & (88\%) \\
2 & $\mathrm{\nu_\mathrm{e} + ^{12}\mathrm{C} \rightarrow ^{12}\mathrm{N} + e^{-}}$ & (17.3 MeV) & (1.5\%)\\ 
3 & $\mathrm{\bar \nu_\mathrm{e} + ^{12}\mathrm{C} \rightarrow ^{12}\mathrm{B} + e^{+}}$ & (14.4 MeV) & (1.0\%)\\
4 & $\mathrm{\nu_\mathrm{i}~ + ^{12}\mathrm{C} \rightarrow \nu_{\mathrm{i}} + ^{12}\mathrm{C}^{*}+ \gamma}$ & (15.1 MeV) & (2.0\%) \\
5 & $\mathrm{\nu_\mathrm{i} + e^{-} \rightarrow \nu_{\mathrm{i}} + e^{-}}$ & (-) & (3.0\%)\\ 
6 & $\mathrm{\nu_\mathrm{e} + ^{56}\mathrm{Fe} \rightarrow ^{56}\mathrm{Co}^{*} + e^{-}}$ & (10. MeV) & (3.0\%)\\
7 & $\mathrm{\bar \nu_\mathrm{e} + ^{56}\mathrm{Fe} \rightarrow ^{56}\mathrm{Mn} + e^{+}}$ & (12.5 MeV) & (0.5\%) \\
8 & $\mathrm{\nu_\mathrm{i}~ + ^{56}\mathrm{Fe} \rightarrow \nu_{\mathrm{i}} + ^{56}\mathrm{Fe}^{*}+ \gamma}$ & (15. MeV) & (2.0\%)\\ 
\enddata
\end{deluxetable}

The Large Volume Detector\footnote{LVD is the successor to the Mont Blanc LSD detector \citep{1992LSD}.} is a 1000 t liquid scintillator experiment aimed at detecting O(MeV) and O(GeV) neutrinos, both of astrophysical origin (like those from supernova explosions) and from accelerators (like those from the CNGS beam, see e.g. \citep{2012Agafonova}). Neutrinos can be detected in LVD through charged current (CC) and neutral current (NC) interactions on proton, Carbon nuclei and electrons of the liquid scintillator. The scintillator detector is supported by an iron structure, whose total mass is about 850 t. This can also act as a target for neutrinos and antineutrinos, as the product of interactions in iron can reach the scintillator and be detected \citep{2007Agafonova}. The total target thus consists of $8.3$ x $10^{31}$ free protons, $4.3$ x $10^{31}$ C nuclei  and $3.39$ x $10^{32}$ electrons in the scintillator and of $9.7$ x $10^{30}$ Fe nuclei in the support structure.
The main neutrino reaction in LVD is the inverse beta decay (IBD), as it can be seen in Table \ref{tab:interactions}, where all other relevant neutrino interaction channels are shown too.

LVD consists of an array of 840 scintillator counters, 1.5 m$^{3}$ each, viewed from the top by three photomultipliers (PMTs). It is a modular detector. From the viewpoint of PMTs power supply, trigger and data acquisition, the array is divided in sectors (dubbed {\it towers}): each sector operates independently of the others. Each tower includes 280 counters, divided in 4 {\it groups} of 80, 80, 64 and 56 counters: they share the same low-voltage power supplies. Moreover, for each tower, counters are organized in 35 {\it modules} of 8 ones that are at the same position in the array. Those share the same charge digitizer board \citep{Bigongiari1990} and the same high voltage divider. This modularity allows LVD to achieve a very high duty cycle, that is essential in the search of unpredictable sporadic events. 
On the one hand, the three independent data acquisition systems, one per tower, minimize (in practice, nullify) the probability of a complete shutdown of the experiment.
On the other hand, failures involving one or more counters do not affect other counters. LVD can thus be serviced during data-taking by stopping only the part of the detector (down to individual counters) that needs maintenance. The modularity of the detector results in a ``dynamic'' active mass $\mathrm{M_{act}}$, as we will see in Section 3.1.

LVD has been in operation since 9 June 1992, its mass increasing from 300 t (about one full ``tower'') to its final one, 1000 t, in January 2001. In the following analysis, we consider data recorded between 9 June 1992 and 31 December 2013. During this period, LVD has been running in two different conditions due to different values of the trigger threshold. The trigger logic (extensively described in \citep{2007Agafonova}) is based on the 3-fold coincidence of the PMTs in a single counters. Given the relevance of the IBD reaction, the trigger has been optimized for the detection of both products of this interaction, namely the positron and the neutron. Each PMT is thus discriminated at two different threshold levels, the higher one, $\mathrm{{\cal E}_H}$,
being also the main trigger condition for the detector array. The lower one ($\mathrm{{\cal E}_L \simeq 0.5}$ MeV) is in turn active only in a 1 ms time-window following the trigger, allowing the detection of $\mathrm{(n,p)}$ captures. Between 9 June 1992 and 31 December 2005 (period {\it P1}) $\mathrm{{\cal E}_H}$ was set to 5 MeV for core counters, i.e., counters not directly exposed to the rock radioactivity (about 47$\%$ of the total) and to 7 MeV for external ones. From 1 January 2006 onwards (period {\it P2}) $\mathrm{{\cal E}_H}$ was set to 4 MeV for all counters, independently of their location. The lower threshold has instead remained constant in both periods. 

Once a trigger is identified, the charge and time of the three summed PMTs signals are stored in a memory buffer. The time is measured with a relative accuracy of 12.5 ns and an absolute one of 100 ns \citep{Bigongiari1990}. One millisecond after the trigger, all memory buffers are read out, independently in the three towers. The mean trigger rate is $\mathrm{\approx 0.005~s^{-1}t^{-1}}$ in period {\it P1} and $\mathrm{\approx 0.013~s^{-1}t^{-1}}$ in period {\it P2}, as shown in table \ref{tab:data_set} together with other features of the two periods of data-taking.

\begin{deluxetable}{ccccccccc}

\tabletypesize{\scriptsize}
\tablecaption{ Data set features in periods {\it P1} and {\it P2}: R$_\mathrm{{tot}}$ is the total trigger rate, R$_{7}$, R$_{10}$, and R$_\mathrm{{L}}$ are the rates of events with energy above 7, 10, 0.5 MeV, respectively, and M$_{\mathrm{act}}$ is the average active mass.\label{tab:data_set}}
\tablewidth{0pt}
\tablehead{\\
\colhead{} & \colhead{$\mathrm{R_{tot}}$} & \colhead{$\mathrm{R_7(E\geq 7MeV)}$} & \colhead{$\mathrm{R_{10}(E \geq 10MeV)}$} & \colhead{$\mathrm{R_L(E\geq 0.5MeV)}$} & \colhead{$\mathrm{M_{act}}$} & \colhead{Exposure} & \colhead{$\mathrm{t_{live}}$} & \colhead{$\mathrm{t_{live}(M_{act}\geq300t})$} \\

\colhead{} &\colhead{$\mathrm{[s^{-1} \cdot t^{-1} \cdot 10^{-4}]}$} &\colhead{$\mathrm{[s^{-1} \cdot t^{-1} \cdot 10^{-4}]}$} & \colhead{$\mathrm{[s^{-1} \cdot t^{-1} \cdot 10^{-4}]}$} & \colhead{$\mathrm{[s^{-1} \cdot t^{-1} \cdot 10^{-4}]}$} & \colhead{[t]} & \colhead{$\mathrm{[t \cdot y]}$} & \colhead{$\mathrm{[days]}$} & \colhead{$\mathrm{[days]}$}\\
}
\startdata
P1 & 50 & 1.4 & 0.28 & $2.4\cdot 10^{6}$ & 576 & 7320 &  4636 & 4419 \\
P2 & 130 & 2.0 & 0.26 & $2.5\cdot 10^{6}$  & 946 & 7560 &  2916 & 2916 \\
\enddata
\end{deluxetable}

\section{Event selection}\label{sec:dataanalysis}

The method used in LVD to search for neutrino bursts from gravitational stellar collapses essentially consists in searching, in the time series of single counter signals (events), for a sequence (cluster) whose probability of being simulated by fluctuations of the counting rate is very low (see Section 4). The higher the event frequency, the higher is the probability of a ``background-cluster', given by accidental coincidences. At the trigger level, the bulk of events in LVD is due to natural radioactivity products both from the rock surrounding the detector and from the material that constitutes the detector itself and to atmospheric muons. The set of cuts described in this section aims at reducing such a background while isolating signals potentially due to neutrinos. The first condition (Section 3.1) functions as a filter to remove events triggered in malfunctioning counters. The second and third conditions (Section 3.2) reject cosmic-ray muons and most of the radioactive background. The fourth one (Section 3.3) refines the rejection of defective counters, through the analysis of the time series of the events. As we will show below, after the background reduction, the counting rate is decreased by a factor of about 400.

\subsection{Counter selection (basic cuts)}\label{sec:dataset}

\begin{figure}[htb]
\includegraphics[angle=0,scale=.40]{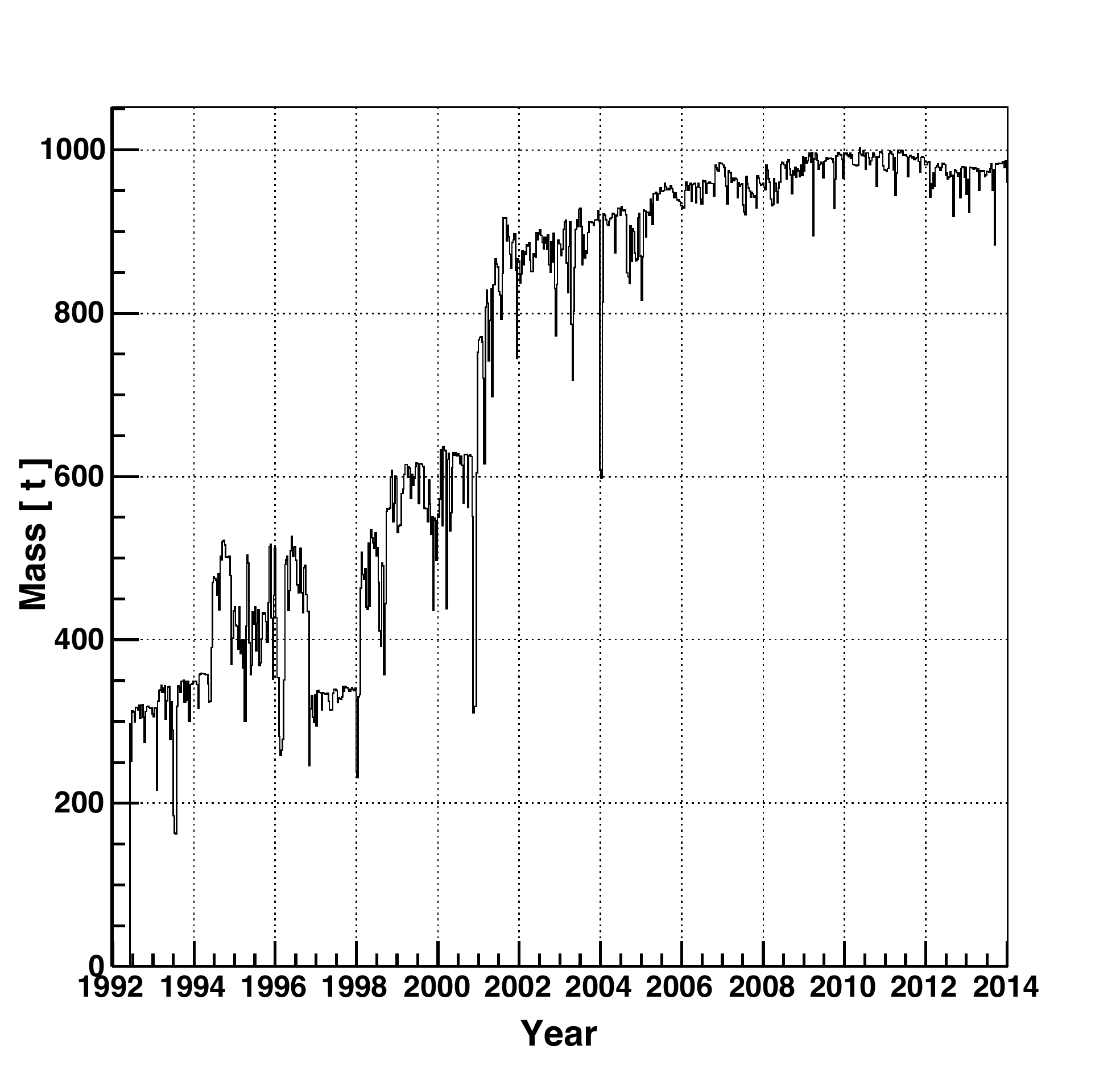}
\caption{LVD active mass as a function of time in the period from June 1992 to December 2013. \label{activemass}}
\end{figure}

The first step in the selection chain is the identification of well-operating counters, i.e. the rejection of signals detected in defective ones. Indeed, the participation of a counter to a trigger does not guarantee its well-functioning. To this aim, we exploit the response of counters to cosmic-ray muons. Muons are identified through the time coincidence of signals in two or more counters. The average rate of muons crossing LVD, $f\mathrm{_{\mu}}$(LVD), is $\mathrm{0.097 \pm 0.010~s^{-1}}$, the measured rate for each counter, $f\mathrm{_\mu}$(c) being $\mathrm{5\cdot10^{-4} s^{-1}}$ (i.e., 1.8/h). The distribution of energy losses of muons in each counter is also monitored: due to the low rate, a muon spectrum is built every month. Quality cuts to be applied to counters (hence to events) are then defined on the basis of muon rate and energy spectra. Namely, we use a counter in the rest of the analysis if $f\mathrm{_\mu(c)\ge 3\cdot 10^{-5}~s^{-1}}$ and if the monthly energy spectrum is consistent with a reference one. Moreover, we require that the counting rate above 7 MeV (corresponding to the high-energy threshold in period {\it P1}) is less than $3 \cdot 10^{-3} \mathrm{s}^{-1}$ during the last two hours of operation \citep{2008Agafonova}. We have verified indeed that high rates usually correspond to faulty electronic or to badly calibrated counters, i.e., counters in need of maintenance. Note that such a cut usually involves a very small amount of counters, 2\% on average.

The active detector mass, M$_\mathrm{{act}}$, resulting after applying the described cuts, is shown in Figure \ref{activemass}, as a function of time in the data period considered in the present analysis. 

\subsection{Neutrino-events selection}
The successive level in the event selection regards the suppression of the muonic and radioactive background. To this aim, the following cuts are applied:\\
- Events characterized by signals in two or more counters within 175 ns are rejected as muons. Furthermore, to avoid  the contamination by any signal associated with muon interactions inside the detector or in the surrounding rock, a dead time of 1 ms is applied after each muon event. The total dead time introduced by this cut is $ t\mathrm{_{dead}\leq 0.01\%}$, corresponding to less than 1 hour per year. The probability of rejecting a neutrino candidate involving more than one counter has been evaluated in \citep{1991Antonioli}. Convolving this probability with the neutrino energy spectra expected from a core collapse supernova we obtain that about 3$\%$ of neutrino interactions are erroneously rejected. Note that in case of a positive neutrino burst identification these events can be recovered. \\
- Only events whose associated energy is in the range $\mathrm{10~MeV \leq E_{signal} \leq 100~MeV}$ are considered. 
This interval is chosen not only with respect to the expected neutrino energy distribution in case of supernova explosions (see for example \citep{2009Pagliaroli_a}) but also because it allows us the suppression of most of the radioactive background. 
In turn, the impact on the expected neutrino signal is small ($\sim 15\%$) because of the energy dependence of neutrino cross sections.

\begin{figure}[htb]
\includegraphics[angle=0,scale=.38]{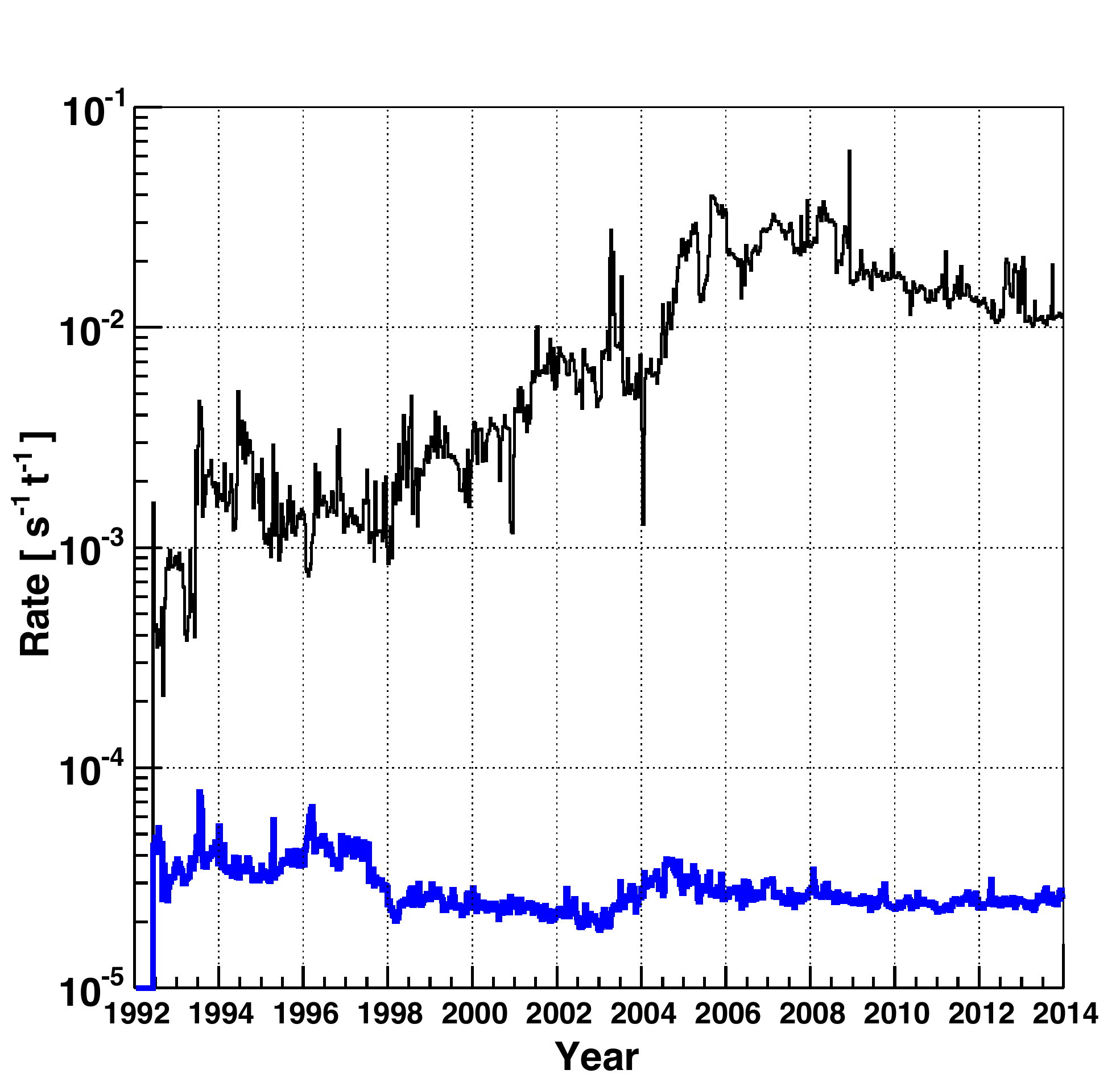}
\caption{LVD counting rate as a function of time in the period from June 1992 to December 2013. The thin black curve shows the trigger rate while the thick blue one shows the rate after the suppression of the muon and radioactive background (see Section 3.2). \label{fig:rate}}
\end{figure}

\begin{figure}[htb]
\includegraphics[angle=0,scale=.40]{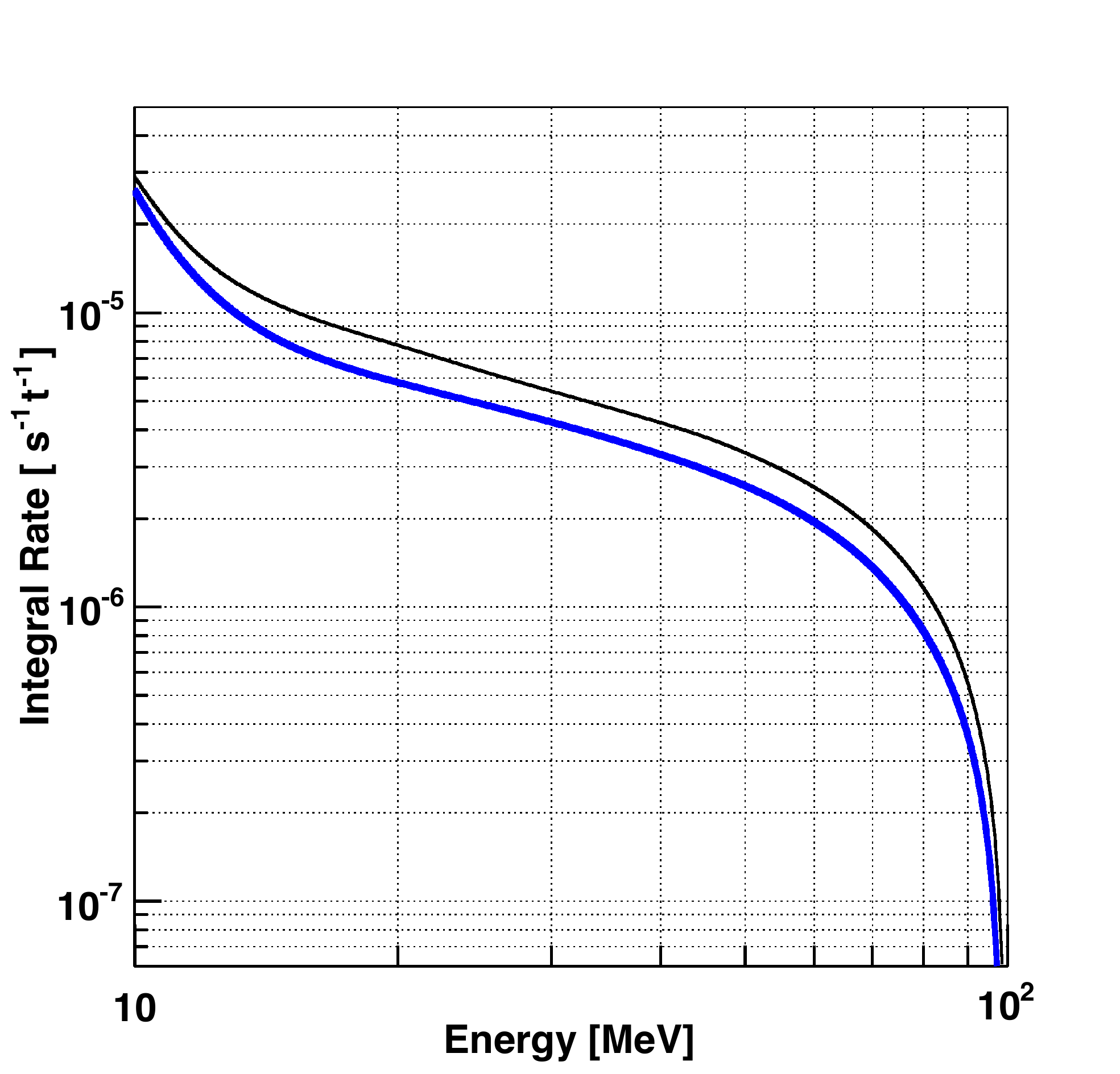}
\caption{Integral energy spectra of signals detected in periods {\it P1} (black thin line) and {\it P2} (blue thick line).}
\label{fig:sint}
\end{figure}

After applying the described cuts, the event rate is strongly reduced
to about $3\cdot 10^{-5}$ s$^{-1}$t$^{-1}$. As it is shown in Figure \ref{fig:rate} it is stable over time and almost independent from the hardware configuration. Indeed, the effect due to the threshold change between the two periods {\it{P1}} and {\it{P2}} is negligible as it is shown in table \ref{tab:data_set} where relevant features of the two periods are listed. Average frequencies $\mathrm{R_{10}}$=$ f\mathrm{(E \geq 10~MeV)}$ and $\mathrm{R_{L}}$ = $f\mathrm{(E \geq 0.5~MeV)}$ are consistent in both periods. The integral energy spectra of the signals, after quality cuts in the energy range $[10-100]$ MeV are also shown, in Figure \ref{fig:sint}, for the two periods. The slightly higher frequency in period {\it P1} is due to the the lower muon discrimination power as a consequence of the smaller active mass. We can conclude that a joint analysis of data taken in the two periods is appropriate. 

\subsection{Counter selection (topological cuts)}\label{sec:dataanalysis-3}
The final set of selection cuts is introduced during the time analysis of the events surviving the previous filters. As we will see in Section 4.2, the search for neutrino bursts in LVD consists in searching for clusters of events within a certain time window, with duration, $\Delta$t, up to 100 s in step of 100 ms. Every cluster is characterized by $\Delta$t and by the number of events, $m$ (multiplicity), within $\Delta$t. In the case of a neutrino burst, we expect the $m$ events to be distributed uniformly over the array, 
thanks to the energy cut at 10 MeV that guarantees the uniformity of the response of counters against threshold effects. If among the $m$ events there is an excess of signals in specific counters, or in 8-counters modules, or in counters groups, then events from that counter (module, group) are rejected.
The following cuts, dubbed ``topological'' as they check the spatial distribution of events, are meant to discard detector components that are temporarily unstable, due for example to electric noise or to maintenance activities in the experiment.

We first check the occurrence of single counters in each cluster of events. If $m$ is the cluster multiplicity and N$\mathrm{_c}$ the number of active counters, a counter is excluded from the cluster if its occurrence n$\mathrm{_c}$ corresponds to a Poisson probability $\mathrm{P(k\geq n_c,~m/N_c) \leq 1 \times 10^{-5}}$. Then we check the occurrence of each 8-counters module in every cluster. The cut on modules is analogous to the one on counters: the probability determining the exclusion is in this case scaled by a factor 8, i.e. $P=8 \times 10^{-5}$. Finally, we apply the same logic as above to check the occurrence of each counters group in every cluster of events. The probability to reject a group is now scaled by a factor 70, i.e. $P=70 \times 10^{-5}$. We note that the sequence of cuts is applied to each cluster of events separately. Also, each cluster is re-analyzed every time a cut (on counters, modules or groups) is applied, i.e., the cluster multiplicity $m$ is re-evaluated at each step of the sequence.

The significance of a cluster (given by its frequency of imitation due to background fluctuations) depends on $\Delta$t and $m$ (see Section 4.2). As ``topological'' cuts might affect (reduce) $m$\footnote{In Section 4.3 we will quantify the effect of these cuts on a possible real neutrino signal.}, we have carefully monitored their incidence on data over time. In particular, we have inspected every cluster that had a rather low imitation frequency (less than 1/month), i.e., rather high significance, before applying topological cuts. Most of them correspond to periods of electronics problems, in particular of time-to-digital converters. A malfunctioning of TDCs can spuriously increase the multiplicity of a cluster: muons are not rejected properly (see Section 3.2) and are identified as neutrino candidates.

\section{Search for neutrino bursts}
\label{search}

In this section we describe the analysis performed on the events satisfying the selection criteria described above. The aim is the search for significant clusters of events that could be indicative of neutrino bursts. The pre-requisite to determine the significance is that the counting rate behavior is Poissonian: this is shown in Section 4.1, by studying the time distribution of the selected events. The search for clusters of events, together with the determination of their statistical significance, is detailed in Section 4.2. The sensitivity of LVD to the detection of neutrino bursts resulting from the described analysis is finally discussed in Section 4.3.

\subsection{Time distribution of the data set}  

\begin{figure}[htb]
\includegraphics[angle=0,scale=.40]{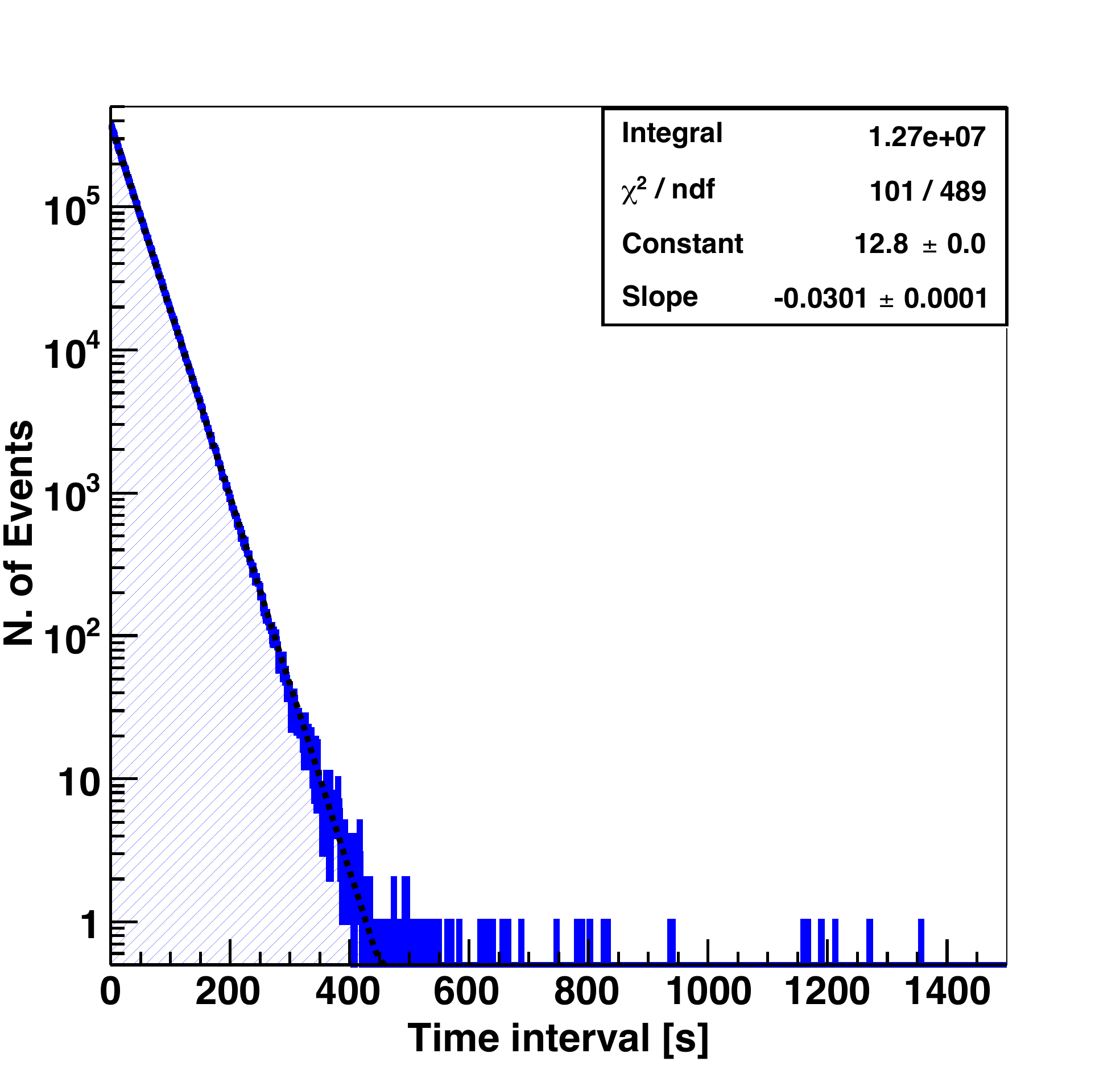}
\caption{Distribution of normalized time intervals between successive events passing the selections described in the text. The normalization is done by equalizing the counting rate at the time of each event to the average one. The dashed black line shows the result of a Poissonian fit to the distribution. \label{Poisson_1}}
\end{figure}

The search for neutrino bursts is performed on data spanning the period from 9 June 1992 to 31 December 2013. During this time the active mass has been larger than 300 t in 7335 days, corresponding to a live time larger than 93\% ($>$ 99 \% since 2001). 300 t is the minimal mass that allows LVD to be sensitive to neutrino bursts over the whole Galaxy (see Section 4.3). The number of events collected in this period and passing the cuts described in the previous section is 12694637. The distribution of time intervals between successive events is shown in Figure \ref{Poisson_1} (blue histogram). Due to the variable detector configuration, the differences in time have been normalized to account for the active mass at the time of the events. The normalization is done by equalizing the event rate ($f$, that depends on the active mass) to a reference one, $f\mathrm{_{ref}}$, that corresponds to the average one when the whole array (1000 t) is in operation, i.e., $\delta t_\mathrm{{norm}}=\delta t \cdot f/f_\mathrm{{ref}}$, with $f\mathrm{_{ref}= 0.03~s^{-1}}$. LVD events behave as a stochastic time series well described by the Poisson statistics as proved by the quality of the fit to a Poisson distribution (shown in the figure as a dashed black line). 

\subsection{Analysis method}
The search for neutrino bursts in LVD data is essentially a two-steps process.

In the first step, we analyze the entire time series\footnote{We choose here not to exploit the capability of LVD to detect both products of the IBD reaction (see Section \ref{sec:dataanalysis}). Indeed, as discussed in \citep{2012Agafonova}, the algorithm applied to all selected events is more sensitive than when applied to events with the IBD signature. It becomes slightly more efficient if we apply it to a mixture of unsigned and signed events, but at the price of loss of simplicity and of independence from models. Finally, by using all events the algorithm is sensible not only to possible neutrino interactions in the liquid scintillator but also in the iron structure \citep{2004Imshennik}.} to search for cluster of events. 
The rationale of the search is that each event could be the first of a possible neutrino burst. As we do not know a priori the duration of the burst, we consider all clusters formed by each event and the n successive ones, with n from 1 to all those contained inside a time window $\mathrm{\Delta t_\mathrm{max}}=100$~s.
The duration of each cluster is given by the time difference $\mathrm{\Delta t}$ between the first event 
and the last one of each sequence. 
The analysis is then applied iteratively, starting from the next one, to all LVD events.
The advantage of the described analysis, where all clusters with durations up to 100 s are considered, is that it is unbiassed with respect to the duration of the possible neutrino burst, unknown a priori. Moreover, the choice of $\mathrm{\Delta t_\mathrm{max}}=100$~s is very conservative as it well exceeds the expected duration of a neutrino burst from core collapse supernovae and even more that from failed supernovae. 

The second step of the process consists in deciding if one or more among the detected clusters are neutrino-bursts candidate. To this aim, we associate to each of them (characterised by $m\mathrm{_{i},\Delta t_{i}}$) a quantity that we call imitation frequency $\mathrm{F_{im_{i}}}$. This represents the frequency with which background fluctuations can produce clusters 
of any duration, between 0 and $\mathrm{\Delta t_{max}}$, with the same or lower probability than that of the individual cluster. As shown in \citep{1996Fulgione}, this quantity, which depends on ($m\mathrm{_{i}, \Delta t_{i}}$), on the background rate, $f\mathrm{_{bk_{i}}}$ and on the maximum cluster duration chosen for the analysis, $\mathrm{\Delta t_{max}}$, can be written as:
\begin{equation}
\mathrm{F_{im_{i}} = }f\mathrm{_{bk}^2 \Delta t_{max} \sum_{k \geq m_{i}-2} P(k,} f\mathrm{_{bk_{i}} \Delta t_{i}) }
\end{equation}

Given the duration of the LVD data set (more than 20 years), we choose 1/100 y$^{-1}$ as imitation-frequency threshold, $\mathrm{F_{im
}^{th}}$. That means that a cluster $(m\mathrm{_{i},\Delta t_{i})}$ is considered as a candidate neutrino burst if:
\begin{equation} 
\sum_{\mathrm{k \geq} m_{\mathrm{i}}-2}\mathrm{P(k,} f_\mathrm{bk} \Delta t_{\mathrm{i}})
< \frac{\mathrm{F^{th}_{im}}}{f^{2}_\mathrm{bk} \cdot \Delta t_\mathrm{max}} 
\end{equation}
where $\mathrm{P(k, f_\mathrm{bk} \Delta t_{\mathrm{i}})}$ is the Poisson probability to have k events in the time window $\mathrm{\Delta t_{\mathrm{i}}}$
if $f\mathrm{_{bk}}$ is the average background frequency.

\begin{figure}[htb]
\includegraphics[angle=0,scale=.40]{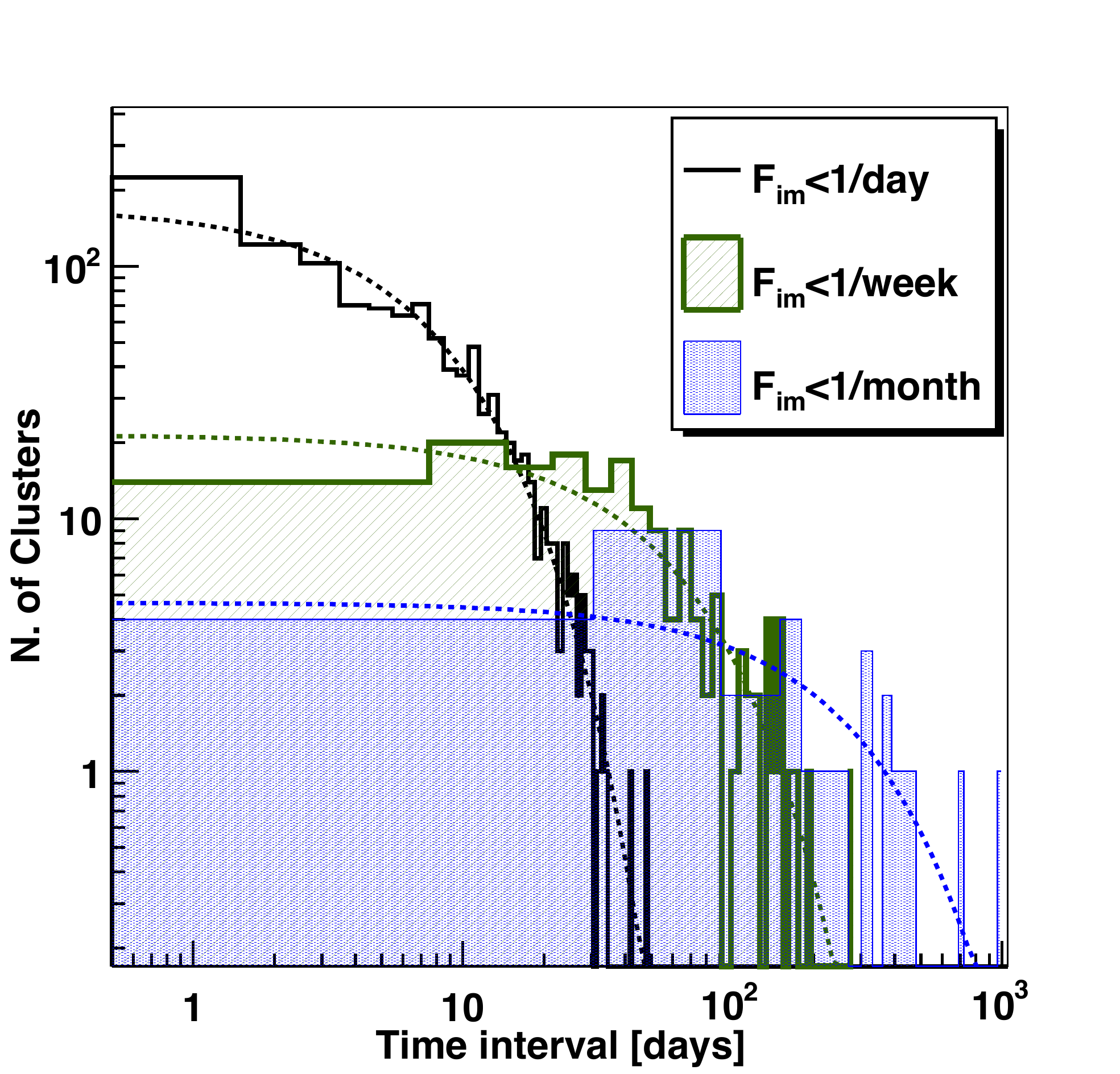}
\caption{Distributions of time intervals between consecutive clusters (solid lines) fitted by Poisson laws (dashed lines) for imitation frequencies: F$\mathrm{_{im} <}$ 1/day (black), 1/week (green) and 1/month (blue). \label{Poisson}}
\end{figure}

\begin{figure}[htb]
\includegraphics[angle=0,scale=.50]{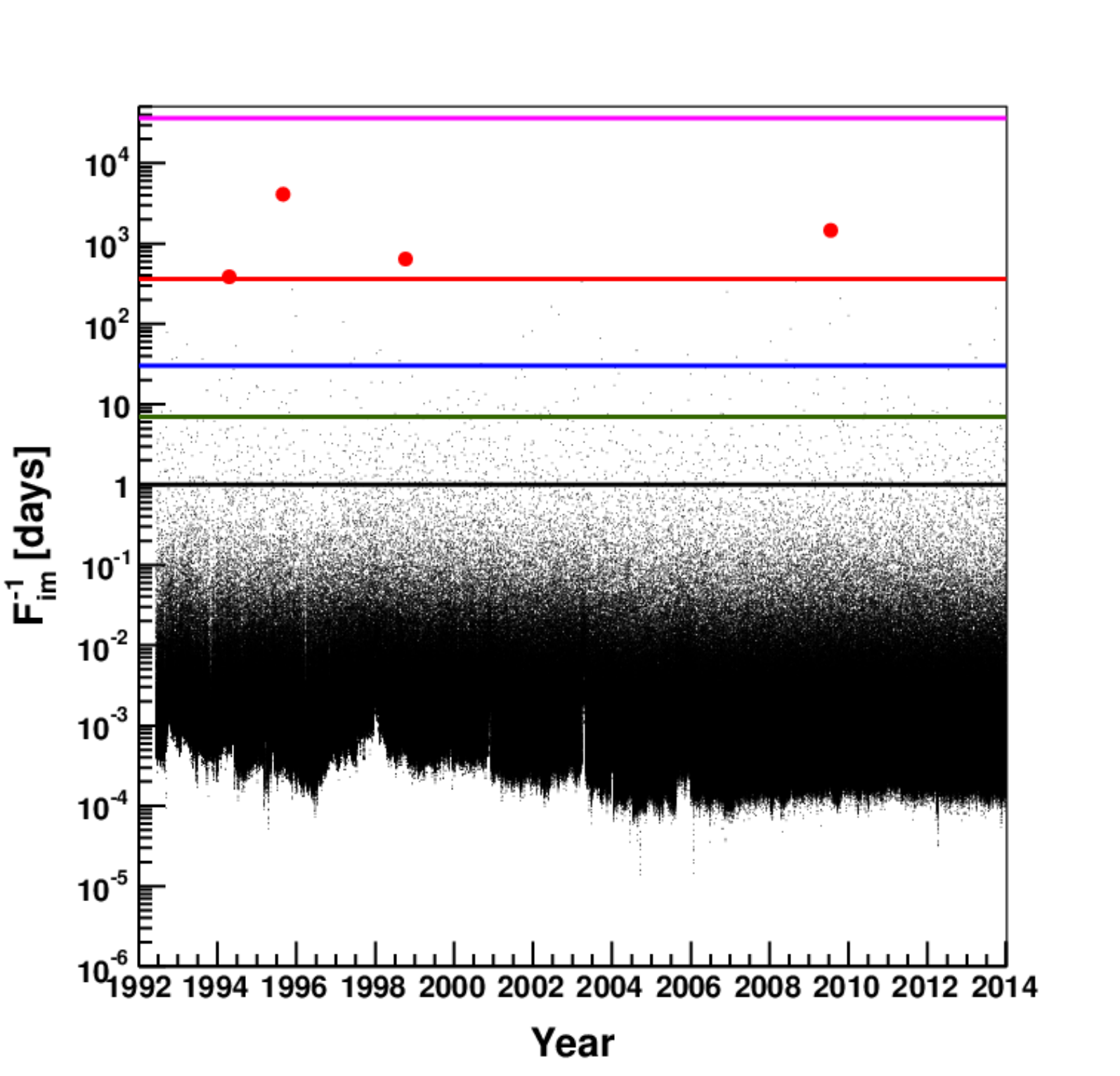}
\caption{Distribution of detected clusters versus time between June 1992 and December 2013. Red dots represent clusters with imitation frequency less than $\mathrm{F_{im} = 1/year}$. Black green, blue, red and purple lines correspond to $\mathrm{F^{th}_{im} = 1/~day}$, $\mathrm{F^{th}_{im} = 1/~week}$, $\mathrm{F^{th}_{im} = 1/~month}$, $\mathrm{F^{th}_{im} = 1/~y}$, $\mathrm{F^{th}_{im} = 1/100~y}$, respectively. \label{allcandidates}}
\end{figure}

The introduction of the imitation frequency has a double advantage. From the viewpoint of the search for neutrino bursts, it allows us to define a priori the statistical ``significance'' of each cluster in terms of frequency. Also, it allows us to monitor the performance of the search algorithm and the stability of the detector\footnote{The performance of the selection procedure and its capability to discriminate a burst from background fluctuations has also been hardware tested, by generating clusters of signals in a subset of counters equipped with a LED system.} by increasing the imitation-frequency threshold. Namely, we study the time distribution of clusters (i.e., the difference in time between clusters) having imitation frequency less than 1/day, 1/week and 1/month. The number of clusters detected during 7335 days are 1123, 165 and 45, respectively. Note that these rates are definitely lower than the corresponding frequency limits (7335, 1048 and 245, respectively). In Figure \ref{Poisson} we show the distributions of time differences between consecutive clusters for the three different values of imitation frequency ($\mathrm{F_{im}}<1$~day$^{-1}$, week$^{-1}$, month$^{-1}$ as black, green, and blue histograms, respectively). The superimposed dotted lines are the result of a Poissonian fit to each distribution. 
The good agreement between data and the expected Poissonian behavior shows that the search algorithm and the detector are under control over the whole period of data taking. Also, the occurrence of clusters with different $\mathrm{F_{im}}$ over 7335 days of measurement is uniform as can be seen in Figure \ref{allcandidates}. It shows the $\mathrm{F_{im}}$ of all detected clusters as a function of time. Clusters above the black, green, blue lines are those with $\mathrm{F_{im}}<1$~day$^{-1}$, week$^{-1}$, month$^{-1}$, respectively. 

\subsection{Analysis sensitivity}

\begin{figure}[htb]
\includegraphics[angle=0,scale=.40]{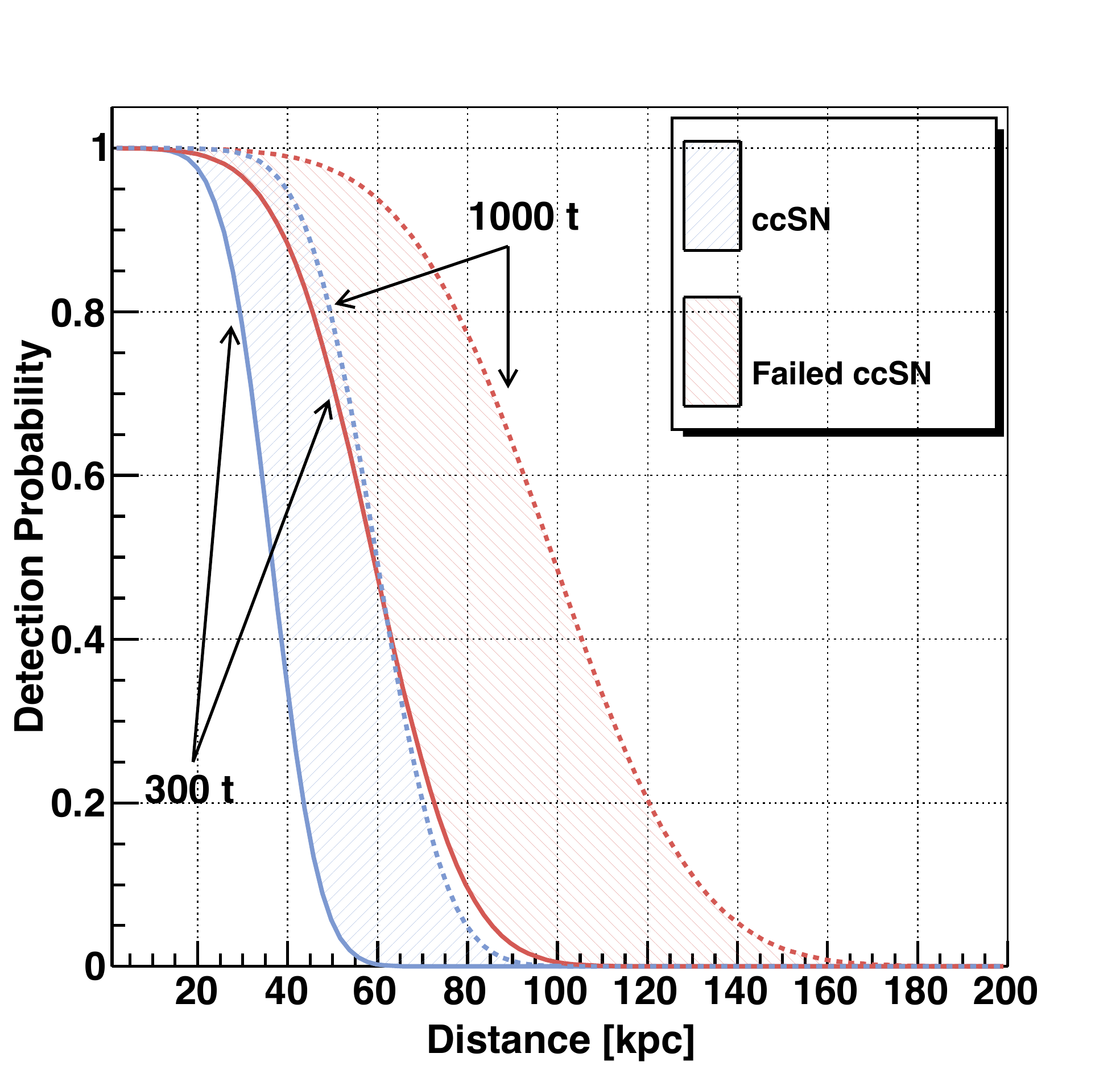}
\caption{LVD detection probability versus source distance for the imitation frequency of 1/100 y$^{-1}$ (see text). The blue and red bands correspond to the case of standard core collapse (ccSN) and failed supernovae, respectively. The solid (dashed) line represents an active mass of 300 (1000) t. \label{eff}}
\end{figure}

The capability of LVD to detect in real-time (i.e., ``on-line'') a supernova event is extensively discussed in \citep{2008Agafonova}. In that case a fixed time-window (i.e., 20 s) is used in the burst-search algorithm. In turn, for the present analysis (so-called ``off-line'') we do not fix a priori the duration of the burst, i.e., we consider all possible durations up to 100 s. Consequently, we extend here our previous study to account for this choice. As in \citep{2008Agafonova}, we discuss the sensitivity to the identification of a neutrino burst in terms of the maximum detectable distance of the supernova explosion.\\
To estimate the characteristics of a neutrino signal in LVD from a gravitational stellar collapse we exploit the parametrization of the neutrino flux proposed by \citep{2009Pagliaroli_a}. That is based on the analysis of neutrinos observed at the occurrence of SN1987A and it includes the impact of neutrino oscillations too. The adopted model can be summarized as follows:\\
- the neutrino emission occurs in two main stages:\footnote{$\nu_{e}$ emitted in shock breakout, when $\nu_{e}$ produced in electron captures (neutronization) are released, play a secondary role in the detector sensitivity.} $\nu_{e}$ and $\bar \nu_{e}$ are emitted during the accretion phase ($\approx$ 500 ms), determining in part the future evolution of the core collapse \citep{2011Ott}; neutrinos and antineutrinos, $\nu_{i}$ and $\bar \nu_{i}$, of all flavors are emitted during the thermal cooling;\\
- the total neutrino signal is expected to develop on a time scale of about 10 s, being 90\% (50\%) the fraction of detected events in the first 10 s (1 s);\\
- the time averaged temperatures of emitted neutrinos are: 10.7 MeV for $\nu_{e}$, 12.0 MeV for $\bar \nu_{e}$ and 14.2 MeV for $\nu_{\mu,\tau},\bar\nu_{\mu,\tau}$;\\
- Mikheev-Smirnov-Wolfenstein (MSW) oscillations effects on neutrinos crossing the matter of the collapsing star \citep{1978Wolfenstein}, \citep{1985Mikheev} are taken into account, while $\nu$-$\nu$ interactions are neglected \citep{2007Agafonova}. The normal mass hierarchy scenario has been conservatively assumed together with the most recent values of $\theta_{12}$ and mass squared differences $\Delta m_{12}^{2}$ and $\Delta m_{23}^{2}$ (see e.g. \citep{updStrumia} for a review). In this scenario the 
non null value for $\theta_{13}$ mixing angle \citep{2012DayaBay} has no significant impact on the expected neutrino signal.\\
By simulating neutrino events in LVD generated according to the described model, we estimate the detection probability as a function of the distance of the gravitational stellar collapse from the Earth. We find that a total of more than 300 events would trigger LVD for a collapse 10 kpc away: events are shared among all interaction channels as shown in Table \ref{tab:interactions}. This number becomes more than 260 taking into account the chosen energy cut at 10 MeV\footnote{As anticipated in Section 3.3, the described simulation allows us to evaluate too the possible impact of topological cuts on a real neutrino burst. It results that even in the worst possible experimental conditions, i.e., for a source at 25 kpc and a minimal detector active mass, $\mathrm{M_{act}=300~t}$, the probability to mistakenly reject counters, modules or groups due to statistical fluctuations of an uniform distribution, thus downgrading an authentic cluster, remains always $< 3 \cdot 10^{-4}$.}. The detection probability as a function of the distance of the collapse is shown in Figure \ref{eff} for the chosen imitation frequency of 1/100 y$^{-1}$. The blue band corresponds to the case of standard core collapse supernovae: the solid (dashed) line represents an active mass of 300 (1000) t.

We also evaluate the detection probability in the case of stellar collapses ending into black-holes, so-called failed supernovae, by using a similar procedure as above. This is shown as a red band in the same Figure \ref{eff} (similarly, the two boundary lines represent a mass of 300 t and 1000 t). In this case, we take as reference the predictions of \citep{2008Nakazato} by choosing the most conservative one in terms of neutrino emission. Namely, we assume a progenitor of 40 solar masses, a burst duration shorter than 500 ms, a total emitted energy in neutrinos of 1.3 10$^{53}$ ergs and the inverted neutrino mass hierarchy.

We can conclude that the LVD efficiency in detecting supernovae or failed supernovae explosions is more than 95\% for distances less then 25 kpc when the detector active mass is larger than $300$ t. 

\section{Results}
\label{results}

\begin{figure}[htb]
\includegraphics[angle=0,scale=.50]{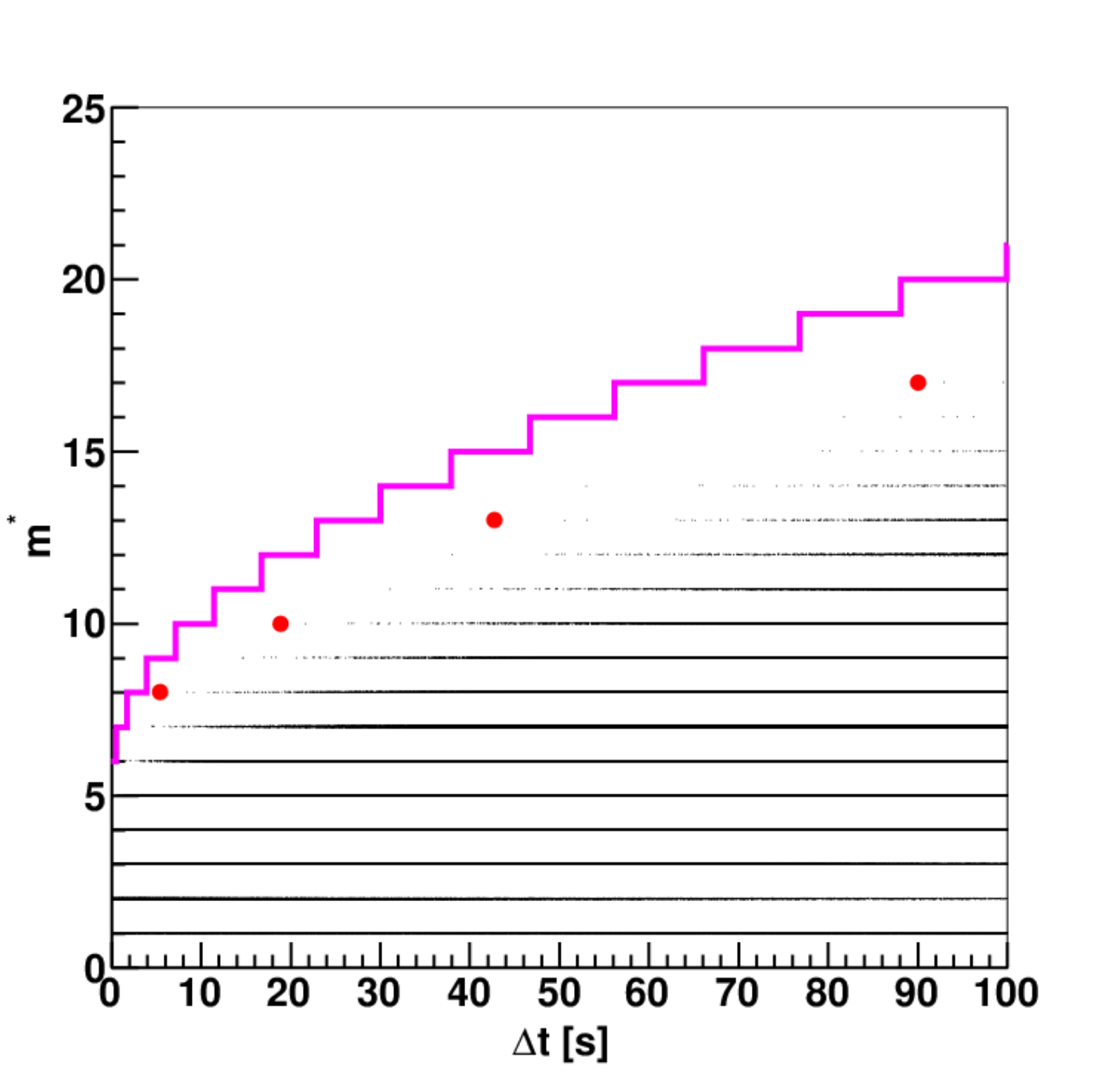}
\caption{Distribution of detected clusters in the space ($\mathrm{\Delta t,m^{*}}$). Red dots represent clusters with imitation frequency less than $\mathrm{F_{im} = 1~y^{-1}}$. The purple line corresponds to $\mathrm{F^{th}_{im} = 1/100~y^{-1}}$. \label{mDeltat}}
\end{figure}

By analyzing the time series of 12694637 events (selected as described in Section 3 and collected over 7335 days of data-taking) we get 26914419 clusters with multiplicity $m\ge2$ and $\mathrm{\Delta t \le 100}$~s.
(12199631 during {\it{P1}} and 14714788 during {\it{P2}}). They are shown in Figure \ref{mDeltat} in a two-dimensional graph whose axes are the cluster duration, $\mathrm{\Delta t}$, and the multiplicity, $m^*$\footnote{$m^*$ is the multiplicity corrected to account for the background frequency, $f_{\mathrm{bk}_\mathrm{i}}$, at the time of each cluster. The correction is done by equalizing $f_{\mathrm{bk}_\mathrm{i}}$ to the average rate $f\mathrm{_{{bk}_{0}}= 0.03~ s^{-1}}$ (see Sections 3.2 and 4.1). $\mathrm{m^*}$ is then obtained by the numerical solution of the equation:
\begin{equation}
\mathrm{\frac{P[k\geq (m^*_i-2), f_{\mathrm{bk}_\mathrm{0}}\cdot \Delta t_i]}{P[k\geq (m_i-2), f_{\mathrm{bk}_\mathrm{i}}\cdot \Delta t_i]} = \frac{f_{\mathrm{bk}_\mathrm{0}}^2}{f_{\mathrm{bk}_\mathrm{i}}^2}}.
\end{equation}
$\mathrm{P(k\geq (m_i-2), f_{\mathrm{bk}_\mathrm{i}}\cdot\Delta t_i)}$ is the Poisson probability to have clusters of multiplicity $\mathrm{k\geq (m_i-2)}$ and $\mathrm{(f_{\mathrm{bk}_\mathrm{i}}\cdot \Delta t_i)}$ is the average multiplicity.}.

\begin{deluxetable}{cc|ccc|ccccc}

\tabletypesize{\scriptsize}
\tablecaption{Characteristics of clusters with significance $\mathrm{F_{im} <1 \cdot \mathrm{y}^{-1}}$: time of occurrence (UTC), active mass (M$_{\mathrm{act}}$), background rate (f$_{\mathrm{bk}}$), distance corresponding to 90\% detection probability (D$_{\mathrm{90\%}}$), multiplicity (m), duration ($\Delta$ t), inverse of imitation frequency (F$_{\mathrm{im}}^{-1}$), average events energy ($\mathrm{\bar E}$), number of IBD candidates (N$_{\mathrm{IBD}}$)
\label{tab:data_alarm}}
\tablewidth{0pt}
\tablehead{
\\
\colhead{n.} & \colhead{UTC} & \colhead{M$\mathrm{_{act} [t]}$} & \colhead{$\mathrm{f_{bk} [s^{-1}]}$} & \colhead{$\mathrm{D_{90\%}[kpc]}$} & \colhead{m} &  \colhead{$\mathrm{\Delta t [s]}$} & \colhead{$\mathrm{F^{-1}_{im}[y]}$} & \colhead{$\mathrm{\bar E~[MeV]}$} & \colhead{$\mathrm{N_{IBD}}$} \\
}
\startdata
1 & 1994 16 April 10:40:49.263 & 346 & $1.08 \cdot 10^{-2}$ & 29.5 & 7 & $18.88$ & $1.06$ & $26.5$ & $2$ \\
2 & 1995 27 August 16:18:10.478 & 431 & $1.85 \cdot 10^{-2}$ & 35.0 & 7 & $5.49$ & $11.16$ & $36.2$ & $1$ \\
3 & 1998 7 October 15:41:41.775 & 552 & $1.40 \cdot 10^{-2}$ & 30.6 & 12 & $90.05$ & $1.76$ & $32.2$ & $3$  \\
4 & 2009 18 July 7:39:20.517& 976 & $2.40 \cdot 10^{-2}$ & 40.4 & 12 & $42.71$ & $4.02$ & $14.6$ & $1$  \\
\enddata
\end{deluxetable}

For each cluster, we evaluate the imitation frequency, $\mathrm{F^{th}_{im}}$, following eq. 1. Those are shown in Figure \ref{allcandidates} as a function of time. In both Figures \ref{allcandidates} and \ref{mDeltat} the purple line represents the expectations for a $\mathrm{F^{th}_{im}}$ of 1/100 y$^{-1}$, i.e., the threshold for considering a cluster as a neutrino-burst candidate (see Section 4.2). None of the observed clusters passes such threshold, the maximum detected significance being $\mathrm{(F_{im})^{-1} = 11.16}$ y associated to a cluster of 7 events during about 5 seconds. 
For the sake of completeness, we have carefully inspected all clusters with F$\mathrm{_{im} \leq}$ 1/month (45 of them). Their energy spectra have been examined as well as the number of low-energy delayed signals that might be the signature of  IBD interactions (see Section 2). All 45 clusters are fully compatible with chance coincidences among background signals. The characteristics of the 4 most significant among them (F$\mathrm{_{im} \leq}$ 1/year) are reported in Table \ref{tab:data_alarm}. Besides the date, we show the conditions of the detector at the time of the cluster, i.e., active mass and background frequency. The properties of the clusters are listed in the last five columns: multiplicity, duration, imitation frequency, average energy of events and number of IBD candidate events. The distance corresponding to 90\% detection probability is also shown: that is derived from the blue curve in Figure \ref{eff} account taken of the active mass. We note that for all four clusters it is well above 25 kpc.

We conclude that no evidence is found for core collapse or failed supernovae during the considered data-taking period. Account taken of the live-time of 7335 days, we obtain a limit on the rate of gravitational collapses out to 25 kpc of less than 0.114 per year at 90\% C.L.

\section{Conclusions}
\label{conclusions}

In this paper we have presented the results of the search for neutrino-burst signals from supernovae explosions performed with LVD data taken over more than 20 years, from 9 June 1992 to 31 December 2013. 

The neutrino-burst detection technique is based on the search for a sequence of candidate neutrino events whose probability of being simulated by fluctuations of the counting rate is very low. As the latter is dominated by the background of atmospheric muons and natural radioactivity products, we have developed a set of selective criteria to isolate signals more probably due to neutrinos. Such a selection is based on the topology and energy of events. Also, given the large number of detectors and the long time of operation, we have been very careful in identifying ill-functioning and/or unstable ones over time. After the selection, the background rate is reduced by a factor of about 400, leaving us with almost 13 millions of events. 

To search for candidate neutrino-bursts among them we have searched for all possible clusters of events with durations up to 100 s. That makes our search model-independent, as the duration of a neutrino burst due to a supernova explosion is unknown. The knowledge of the background as well as its long-term stability are of essence for evaluating the probability of each found cluster. We have shown that the counting rate is stable over the period of observation and that its behavior is Poissonian. That has allowed us to associate to each candidate burst an a-priori significance, that we have chosen to give in terms of imitation frequency, F$_\mathrm{im}$. Given the total time of observation, we have fixed a threshold to F$^{\mathrm{th}}_\mathrm{im}$ of 1/100 y$^{-1}$ for considering a cluster as a real neutrino-burst. We have shown that with the adopted method of analysis and with the chosen threshold, LVD is fully efficient to gravitational collapses (due to supernovae explosions or failed supernovae) within a radius of 25 kpc from Earth, even when its mass is only one third (300 t) of its full one (1000 t). 

Out of the 27 millions of detected clusters, we have found that none has an imitation frequency less than 1/100 y$^{-1}$. We have thus concluded that no evidence has been found for core-collapse supernovae occurred up to 25 kpc during the period of observation\footnote{During this time, results of the search have been communicated every two years at International Cosmic Ray Conferences \citep{ICRC}}. Finally we have set a limit of less than 0.114 collapses per year at 90\% C.L., this being the most stringent limit ever achieved by the observation of supernovae through neutrinos in the entire Galaxy.

\section{Acknowledgments}
The authors wish to thank all 
the staff of the Gran Sasso National Laboratory for
their constant support and cooperation during all these years. The successful installation, commissioning, and operation of LVD would not have been possible without the commitment and assistance of the technical staff of all LVD institutions.
We are grateful to Francesco Vissani for innumerable discussions and clarifications on various physical aspects of the core collapse supernova problem.
One of the author (W.F.) is indebted with Pio Picchi for the idea to express the event significance in terms of background frequency, with Giulia Pagliaroli for the suggestion to expand the interpretation of our results to failed supernovae and with Marco Grassi and the Gran Sasso Scientific Committee for constructive discussions.   
Finally, some of the scientists who imagined, realized and contributed to the LVD experiment are not with us anymore. We wish to remember here in particular Carlo Castagnoli, Gianni Navarra and Georgiy T. Zatsepin: we are left with their memory and their teachings.

%


\begin{thebibliography}{}
\bibitem[Adams et al.(2013)]{2013Adams} Adams Scott M. et al., 2013, ApJ, 778:164 
\bibitem[Agafonova et al.(2007)]{2007Agafonova} Agafonova N.Yu. et al., 2007, APh, 27, 254-270; [hep-ph/069305].
\bibitem[Agafonova et al.(2008)]{2008Agafonova} Agafonova N.Yu. et al., 2008, APh, 28, 516
\bibitem[Agafonova et al.(2012)]{2012Agafonova} Agafonova N.Yu. et al., 2012, PhRvL, 109-7, 070801
\bibitem[Aglietta et al.(1987)]{1987Aglietta} Aglietta, M. et al., 1987, EPL, 3, 1315
\bibitem[Aglietta et al.(1992)]{1992Aglietta} Aglietta, M. et al., 1992, NcimA, 105, 1793
\bibitem[LSD Collaboration(1992)]{1992LSD} M.Aglietta {\it et al.}, 1992, APh, 1, 1-9.
\bibitem[Ahrens et al.(2002)]{2002Ahrens} Ahrens, J., et al. 2002, APh, 16, 345
\bibitem[Aharmim et al.(2011)]{2011Aharmim} Aharmim B. et al., 2011, ApJ, 728:83
\bibitem[Alekseev et al.(1987)]{1987Alekseev} Alekseev, E. N., et al., 1987, JETPL, 45, 589
\bibitem[Ambrosio et al.(2004)]{2004Ambrosio} Ambrosio, M., et al. 2004, EPJC, 37, 265
\bibitem[An et al.(2012)]{2012DayaBay} An, F. P., et al., 2012, PhRvL, 108, 171803
\bibitem [Antonioli et al.(1991)]{1991Antonioli} P. Antonioli, W. Fulgione, P. Galeotti and L. Panaro, 1991, NIMPA, 309, 569.
\bibitem[Antonioli et al.(2004)]{2004Antonioli} Antonioli, P., et al., 2004, NJPh, 6, 114
\bibitem[Bahcall et al.(1995)]{1995Bahcall} Bahcall, J.N., Kamionkowski M. and Sirlin, A., 1995, PhRvD, 51, 6146
\bibitem[Bettini (1999)]{1999Bettini} Bettini, A. 1999. The Gran Sasso Laboratory 1979-1999: a vision becomes reality. Assergi: Laboratori Nazionali del Gran Sasso. 
\bibitem[Bigongiari et al.(1990)]{Bigongiari1990} Bigongiari A., Fulgione W., Passuello D., Saavedra O., Trinchero G., 1990, NIMPA, 288,  529.
\bibitem[Bionta et al.(1987)]{1987Bionta} Bionta, R. M., et al. (IMB collaboration), 1987, PhRvL, 58, 1494
\bibitem[Colgate and White(1966)]{1966Colgate} Colgate S.A. and White R.H., 1996, ApJ, 143, 626
\bibitem[Domogatsky and Zatsepin(1965)]{1965Zaz} Domogatsky G.V. and Zatsepin G.T., 1965, in 9th ICRC Conf. Proc., Vol.1 1030  
\bibitem[Falk(1978)]{1978Falk} Falk S.W., 1978, ApJ, 225, L133
\bibitem[Fukugita et al.(1988)]{1988Fukugita} Fukugita M., Kohyama Y., Kubodera K.,1988, PhLB, 212, 139
\bibitem[Fulgione et al.(1996)]{1996Fulgione} Fulgione W., Mengotti-Silva N., and Panaro L., 1996, NIMPA, 368, 512
\bibitem[Gamow and Shoenberg(1940)]{1940Gamow} Gamow G. and Shoenberg M., 1940, PhRv, 58, 1117
\bibitem[Hirata et al.(1987)]{1987Hirata} Hirata, K., et al., 1987, PhRvL, 58, 1490
\bibitem[Ikeda et al.(2007)]{2007SuperK}Ikeda, M., et al., 2007, ApJ, 669, 519
\bibitem[Imshennik \& Ryazhskaya(2004)]{2004Imshennik} Imshennik, V.S. \& Ryazhskaya, O., 2004, AstL, 30, 14 
\bibitem[Loredo and Lamb(2002)]{2002Loredo} T.J. Loredo, D.Q. Lamb, 2002, PhRvD, 65, 063002
\bibitem[LVD Collaboration(1993-2013)]{ICRC} LVD Collaboration, 1993-2013, in ICRC Conf. Proc.
\bibitem[Klein and Chevalier(1978)]{1978Klein} Klein R.I. and Chevaliere R.A. 1978, ApJL, 223, L109
\bibitem[Kolbe and Langanke(2001)]{2001Kolbe} Kolbe E., Langanke K., 2001, PhRvC, 63, 025802
\bibitem[Mikheev and Smirnov(1985)]{1985Mikheev} Mikheev, S. P., \& Smirnov, A. Y. 1985, SvJNP, 42, 913
\bibitem[Nadyozhin(1977)]{1977Nadyozhin} Nadyozhin D.K., 1977, ApSS, 49, 399 and 1978, 53,131
\bibitem[Nakazato et al.(2008)]{2008Nakazato} Nakazato K., Sumiyoshi K., Suzuki H. and Yamada S., 2008, PhRvD, 78, 083014
\bibitem[Novoseltseva et al.(2011)]{2011Novoseltseva} Novoseltseva, R. V., Boliev, M. M., Volchenko, V. I., et al. 2011, Proc. 32nd ICRC (Beijing), 4, 153 
\bibitem[O'Connor and Ott(2011)]{2011Ott} E. OÕConnor and C. D. Ott, 2011, ApJ, 730, 70
\bibitem[Pagliaroli, Vissani, Costantini and Ianni(2009)]{2009Pagliaroli_a} Pagliaroli G., Vissani F., Costantini M.L., Ianni A., 2009, AsPh, 31, 163
\bibitem[Pagliaroli, Vissani, Coccia and Fulgione(2009)]{2009Pagliaroli_b} Pagliaroli G., Vissani F., Coccia E., Fulgione W. 2009, PhRvL, 103, 031102
\bibitem[Strumia and Vissani(2010)]{updStrumia} Strumia A. and Vissani F., arXiv:hep-ph/0606054v3
\bibitem[Toivanen et al.(2001)]{2001Toivanen} Toivanen J., Kolbe E., Langanke K., Martinez-Pinedo G. and Vogel P., 2001, NuPhA, 694, 395 
\bibitem[Wolfenstein(1978)]{1978Wolfenstein} Wolfenstein, L. 1978, PhRvD, 17, 2369
\bibitem[Woosley \& Janka(2005)]{Woosley 2005} Woosley S. and Janka T., 2005, NatPh 1, 147-154 
\bibitem[Zel'dovich and Guseinov(1965)]{1965Zel'dovich} Zel'dovich, Ya. B., and Guseinov, O. Kh. 1965, SPhD, 10, 524 
\bibitem[Zichichi (2000)]{2000Zichichi} Zichichi, A. 2000, Subnuclear Physics in World Scientific Series in 20th Century Physics, Vol. 24, The First 50 Years: Highlights from Erice to ELN, ed. by O.Barnabei, P.Pupillo, and F.Roversi Monaco (Singapore: World Scientific)
\end{thebibliography}
\end{document}